\documentclass[12pt,aps,showpacs,floatfix,onecolumn,tightenlines]{revtex4}
\usepackage{epsfig}
\usepackage{psfig}
\usepackage{graphicx}
\usepackage{graphics}
\usepackage{xspace}
\usepackage{amssymb}
\usepackage{amsmath}
\usepackage{latexsym}
\usepackage{natbib}
\usepackage{mathrsfs}
\newcommand{\inieq}{\begin{eqnarray}}            %inizio formula numerata
\newcommand{\fineq}{\end{eqnarray}}            %fine formula numerata 
\newcommand{\diff}{{\rm\,d}}                    %simbolo di derivata totale
\def\p{\mbox{\boldmath $p$}}

\def\q{\mbox{\boldmath $q$}}

\def\k{\mbox{\boldmath $k$}}

\begin{document}
%\begin{frontmatter}
\title{Neutrino-nucleus quasi-elastic scattering and strange quark effects} 

\author{Andrea Meucci} 
\author{Carlotta Giusti}
\author{Franco Davide Pacati }
\affiliation{Dipartimento di Fisica Nucleare e Teorica, 
Universit\`{a} degli Studi di Pavia and \\
Istituto Nazionale di Fisica Nucleare, 
Sezione di Pavia, I-27100 Pavia, Italy}

%\date{\today}

\begin{abstract}
A relativistic distorted-wave impulse-approximation model is applied to 
quasi-elastic neutrino-nucleus scattering.
The bound state is obtained in the framework of the relativistic mean
field theory and the final state interaction is taken into account
in the scattering state. Results for the charged- and
neutral-current neutrino (antineutrino) reactions on $^{12}$C target nucleus
are presented and the sensitivity of various quantities to
the strange quark content of the nucleon weak current is discussed. 

\end{abstract}

\maketitle

\section{Introduction}

Neutrino physics has gained in recent years a wide interest that goes beyond the
study of its intrinsic properties and extends to different fields, such as
cosmology, astro, nuclear, and particle physics. The energy production in the
sun, $\nu$-nucleosynthesis, the dynamics of a supernova process, and possible
CP-violations in neutrino oscillations are just a few examples of problems
involving neutrinos in astrophysics and cosmology \cite{astro}. In hadronic 
and nuclear
physics neutrinos can give invaluable information about the structure of the
hadronic weak neutral-current and the role of the strange quark contribution to
the spin structure of the nucleon. In order to investigate these fundamental
questions various reactions have been proposed: deep inelastic scattering of
neutrinos or of polarized charged leptons on protons \cite{baz,Adams:1997tq},
pseudoscalar meson scattering on protons \cite{Dover:1990ic}, and
parity-violating (PV) electron scattering \cite{fein,wale,mck,beck,nap}. 

A number of PV electron scattering measurements have been carried out in recent
years. First results from the SAMPLE experiment \cite{mul} at the MIT-Bates 
Laboratory and the HAPPEX collaboration \cite{aniol} at Jefferson Laboratory
(JLab)  
seemed to indicate a relatively small strangeness contribution to the proton 
magnetic moment \cite{spa,has} and that
the strange form factors must rapidly fall off at large $Q^2$, 
if the strangeness radius is large \cite{aniol}. 
The SAMPLE collaboration has recently reported \cite{spa2} a new determination 
of the strange quark contribution to the proton magnetic form factor using a 
revised analysis of data in combination with the axial form factor of the proton
\cite{ito}.  
The HAPPEX2 collaboration \cite{hap2} and the G0 experiment \cite{g0} at
JLab, the A4 collaboration \cite{a4} at Mainz 
Microtron, and the E158 experiment \cite{Anthony:2005pm} at SLAC aim at exploring 
the strangeness contribution
through improved measurements over a larger range of momentum transfer. First
results from HAPPEX2 \cite{happex2} on elastic $ep$ scattering and on 
elastic scattering of polarized
electrons from $^4$He 
yield a value of the electric strange form factor of the
nucleon consistent with zero while the magnetic strange form factor seems to 
prefer positive value though zero is still compatible with the data.
Another experiment at JLab \cite{he4p} plans to measure the PV 
asymmetry using $^{208}$Pb as target nucleus. 
A recent review 
of the present situation and a discussion of the theoretical perspectives
of this topic can be found in Ref. \cite{ram}.

Neutrino reactions are a well-established alternative to PV electron scattering
and give us complementary information about the contributions of the sea quarks
to the properties of the nucleon. While PV electron scattering  is essentially
sensitive to the electric and magnetic strangeness of the nucleon,
neutrino-induced reactions are primarily sensitive to the axial-vector form
factor of the nucleon.
A measurement of $\nu (\bar\nu$)-proton elastic scattering at Brookhaven
National Laboratory (BNL) \cite{bnl} suggested a non-zero value for the strange 
axial-vector form factor of the 
nucleon. However, it has been shown in Ref. \cite{gar} that the BNL data 
cannot provide us
decisive conclusions about the strange form factors when also strange vector 
form factors are taken into account.
The BooNE experiment \cite{mini} at Fermi National Laboratory (FermiLab) 
aims,
through the FermiLab Intense Neutrino Scattering Scintillator Experiment 
(FINeSSE) \cite{fin}, 
at performing a detailed investigation of the strangeness contribution to the 
proton spin $\Delta {\mathrm s}$ via neutral-current elastic scattering. 
A determination of the strange form factors through a combined analysis of 
$\nu p$, $\bar \nu p$, and $\vec e p$ elastic scattering is performed 
in Ref. \cite{pate}.

Since an absolute cross section measurement is a very hard experimental task due
to difficulties in the determination of the neutrino flux, in Ref. \cite{gar93}
the measurement of the ratio of proton-to-neutron cross sections in 
neutral-current neutrino-nucleus scattering was proposed as
an alternative way to extract $\Delta {\mathrm s}$. This ratio is sensitive 
to $\Delta {\mathrm s}$ but it is difficult to measure with the desired 
accuracy because
of difficulties in the neutron detection. Thus, the FINeSSE experiment will
focus on the neutral- to charged-current ratio \cite{fin}. Since a significant
part of the events at FINeSSE will be from $^{12}$C, nuclear structure effects 
have to be clearly understood in order to reach a reliable interpretation of
neutrino data. 

At intermediate energy, quasi-elastic
electron scattering calculations \cite{book}, which were able to successfully 
describe a
wide number of experimental data, can provide us a useful tool to study
neutrino-induced processes. However, a careful analysis of neutral-current 
neutrino-nucleus
reactions introduces additional complications, as the final neutrino cannot
be measured in practice and a final hadron has to be detected: 
the corresponding cross
sections are therefore semi-exclusive in the hadronic sector and inclusive in 
the leptonic one. 

General review papers about neutrino-nucleus interactions can be found in Refs.
\cite{Walecka,peccei,alb,kolbe03}.
Both weak neutral-current (NC) and charged-current (CC) scattering have 
stimulated detailed analyses in the intermediate-energy region and  
different approaches have been applied to investigate such processes.  
The CC reaction was studied in Refs.
\cite{min,engel,kim,singh,hayes,volpe,alberico,kolbe,
jacho,kim96,kim96a,umino,co1} using a variety of methods including, for example,
Fermi gas, random phase approximation (RPA) and shell model calculations. 
%The neutral-current
%neutrino-induced knockout from nuclei in Ref. \cite{hor}, where binding energies
%corrections and strange-quark axial-vector and vector effects were investigated.
Detailed analyzes of nuclear structure effects on the determination of
strangeness contribution in neutral-current
neutrino-nucleus scattering were performed in Refs. \cite{alberico,barbaro},
where the relativistic Fermi gas (RFG) model 
%a {\lq\lq hybrid\rq\rq} model in which bound nucleons are
%described by harmonic oscillator wave-functions 
as well as a relativistic 
shell model including final state interactions (FSI) were used. The effects 
of FSI were also studied in Ref. \cite{bleve} within the RFG model, 
in Ref. \cite{gar93} in the framework of RPA, and in 
Ref. \cite{Botrugno:2005kn} in the  
continuum RPA (CRPA) theory. A CRPA model was developed in 
Ref. \cite{ryc}, where 
nucleosynthesis processes were also discussed. 

The sensitivity of the neutrino-nucleus 
cross sections to $\Delta {\mathrm s}$ was also examined in a 
relativistic plane wave impulse approximation (RPWIA) in 
Ref. \cite{vdv}, where a model-independent
approach based on eight structure functions was developed and applied to both
NC and CC reactions. The effects of FSI are generally large 
on the cross section, but they are usually reduced when studying ratios of 
cross sections. This was discussed for the ratio of proton-to-neutron cross 
sections in neutral-current scattering in Refs. \cite{alberico,hor,gar92}. 
In Ref. \cite{Martinez:2005xe} two relativistic models where FSI are treated 
with an optical potential and with a multiple-scattering Glauber approximation 
were presented and compared. At lower energies the optical potential approach 
is preferable, whereas at higher energies the Glauber approximation appears more
natural. Within RPWIA the two models in Ref. \cite{Martinez:2005xe} give 
nearly identical results. When FSI are included the Glauber approach yields
valid results down to nucleon kinetic energies of 200 MeV. The same Glauber
approximation was used in Ref.~\cite{Lava:2005pb} to discuss the sensitivity 
to $\Delta {\mathrm s}$ of the helicity asymmetry of the ejected nucleon.

In this paper we present a relativistic distorted-wave impulse-approximation
(RDWIA) calculation of CC and NC
$\nu$- and $\bar\nu$-nucleus reactions 
in the quasi-elastic region, where the 
neutrino interacts with only one nucleon in the nucleus and the other 
nucleons remain spectators. The CC inclusive $\nu$-nucleus scattering was 
described in Ref. \cite{cc} through a relativistic Green's function approach, 
that was firstly applied to the inclusive quasi-elastic electron scattering in 
Ref. \cite{ee} and where FSI are accounted for by means of a complex optical
potential but without loss of flux. In this paper, the CC $\nu$-nucleus 
reaction is considered semi-exclusive in the hadronic sector and is treated
within the same RDWIA approach that was successfully applied 
in Refs.~\cite{meucci1,meucci2,meucci3,meucci4} to electromagnetic one-nucleon 
knockout reactions. The NC
$\nu$-nucleus process was already studied in Ref. \cite{nc}. Here, we present a 
revised version of our calculations and discuss the effects of
possible strange contributions to various observables.
%The relativistic 
%bound state wave-functions are solutions of a Dirac equation 
%containing scalar and vector potentials obtained in the framework of the 
%relativistic mean field theory \cite{adfx,lala}.
%The FSI are taken into account through a relativistic
%optical model potential \cite{chc,clark}.
%The effective Pauli reduction has been adopted for the outgoing nucleon
%wave-function and the
%resulting Schr\"odinger-like equations are solved for each partial wave.

The formalism is outlined in Sec. \ref{sec.for}. Results are presented and
discussed in Sec. \ref{results}. Some conclusions are drawn in 
Sec. \ref{conc}.

\section{The formalism for the semi-exclusive quasi-elastic scattering}
\label{sec.for}

The $\nu$($\bar\nu$)-nucleus cross section for the
semi-exclusive process can be 
written as a contraction between the lepton and the hadron 
tensor, i.e.,
\begin{eqnarray}
\diff \sigma = \frac {G_{\mathrm{F}}^2 } {2} \ 2\pi \
 L^{\mu\nu}\ W_{\mu\nu}\ \frac {\diff^3k} {(2\pi)^3} \ 
 \frac {\diff^3p_{\mathrm N}} {(2\pi)^3} \ ,
\label{eq.cs1}
\end{eqnarray}
where $G_{\mathrm{F}} \simeq 1.16639 \times 10^{-11}$ MeV$^{-2}$ is the Fermi 
constant,  
$k^\mu_i = (\varepsilon_i,\k_i)$, $k^\mu = (\varepsilon,\k)$ are 
the four-momentum of the incident and final leptons,
respectively, and $\p_{\mathrm N}$ is the momentum of the emitted nucleon. 
For charged-current processes $G_{\mathrm{F}}^2$ has to be multiplied 
by $\cos ^2\vartheta_{\mathrm C} \simeq 0.9749$, where 
$\vartheta_{\mathrm C}$ is the Cabibbo angle.

Here, we assume the reference frame where the $z$-axis is parallel to the 
momentum transfer $\q = \k_i - \k$ and the $y$-axis is parallel to 
$\k_i \times \k$. 

The lepton tensor is defined in a similar way as in electromagnetic knockout 
and can be
written as in Refs. \cite{book,cc,nc}. After projecting 
into the initial and the final lepton 
states, it separates into a 
symmetrical and an antisymmetrical component, i.e.,
\begin{eqnarray}
L^{\mu\nu} = \frac {2} {\varepsilon_i \varepsilon} 
\left[ l_S^{\mu\nu} \mp l_A^{\mu\nu} \right],
\label{eq.lt2}
\end{eqnarray}
with
\begin{eqnarray}
l_S^{\mu\nu} &=& k_i^\mu \ k^\nu + k_i^\nu \ k^\mu - g^{\mu\nu} \ k_i \cdot k
\nonumber \\
l_A^{\mu\nu} &=& i \ \epsilon ^{\mu\nu\alpha\beta} k_{i\alpha} k_\beta ,
\label{eq.lt3}
\end{eqnarray}
where $\epsilon ^{\mu\nu\alpha\beta}$ is the antisymmetric tensor with 
$\epsilon_{0123} = - \epsilon^{0123} = 1$.
The upper (lower) sign in Eq. \ref{eq.lt2} refers to $\nu$($\bar\nu$) 
scattering.

The hadron tensor is given in its general 
form by suitable bilinear products of the 
transition matrix elements of the nuclear weak-current operator 
$J^{\mu}$ between
the initial state $|\Psi_0\rangle$ of the nucleus, of energy $E_0$, and the 
final states of energy $E_{\textrm {f}}$, that are
given by the product of a discrete (or continuum) state $|n\rangle$ of the 
residual
nucleus and a scattering state $\chi^{(-)}_{\p_{\mathrm N}}$ of the emitted 
nucleon, with momentum $\p_{\mathrm N}$ and energy $E_{\mathrm N}$. One has
\begin{eqnarray}
W^{\mu\nu}(\omega,q) & = & 
 \sum_{n}  \langle n;\chi^{(-)}_{\p_{\mathrm N}} 
\mid J^{\mu}(\q) \mid \Psi_0\rangle 
\langle 
\Psi_0\mid J^{\nu\dagger}(\q) \mid n;\chi^{(-)}_{\p_{\mathrm N}}\rangle 
\nonumber \\
& \times & \delta (E_0 +\omega - E_{\textrm {f}}) \ ,
\label{eq.ha1}
\end{eqnarray}
where the sum runs over all the states of the residual nucleus.  
In the first
order perturbation theory and using the impulse approximation, 
%the
%incident neutrino interacts with only one nucleon, while the other ones behave 
%as spectators. Thus, 
the transition amplitude is assumed to be adequately 
described as the sum of terms similar to those appearing in the electron
scattering case \cite{book,meucci1}
\begin{eqnarray}
\langle 
n;\chi^{(-)}_{\p_{\mathrm N}}\mid J^{\mu}(\q) \mid \Psi_0\rangle 
= 
\langle\chi^{(-)}_{\p_{\mathrm N}}\mid   j^{\mu}
(\q)\mid \varphi_n \rangle  \ , \label{eq.amp}
\end{eqnarray} 
where 
$\varphi_n = \langle n | \Psi_0\rangle$
describes the overlap between the initial nuclear state and
the final state of the residual nucleus, corresponding to one hole in the 
ground state of the target. 
The single-particle current operator related to the weak current is  
\begin{eqnarray}
  j^{\mu} &=&  F_1^{\textrm V}(Q^2) \gamma ^{\mu} + 
             i\frac {\kappa}{2M} F_2^{\textrm V}(Q^2)\sigma^{\mu\nu}q_{\nu}	  
	     -G_{\textrm A}(Q^2)\gamma ^{\mu}\gamma ^{5}     
%+ F_{\textrm P}(Q^2)q^{\mu}\gamma ^{5}
 \ \ ({\mathrm {NC}}) 
\ , \nonumber 
	     \\  	     
  j^{\mu} &=&  \Big[F_1^{\textrm V}(Q^2) \gamma ^{\mu} + 
             i\frac {\kappa}{2M} F_2^{\textrm
	     V}(Q^2)\sigma^{\mu\nu}q_{\nu}	  
\nonumber \\  &-&G_{\textrm A}(Q^2)\gamma ^{\mu}\gamma ^{5} + 
 F_{\textrm P}(Q^2)q^{\mu}\gamma ^{5}\Big]\tau^{\pm} \  \ ({\mathrm {CC}}) \ ,  
	     \label{eq.nc}
\end{eqnarray}
where $\tau^{\pm}$ are the isospin operators, $\kappa$ is the anomalous part of 
the magnetic moment, $q^{\mu} = (\omega , \q)$ with $Q^2 = |\q|^2 - \omega^2$ 
is the four-momentum transfer, and
$\sigma^{\mu\nu}=\left(i/2\right)\left[\gamma^{\mu},\gamma^{\nu}\right]$.
$G_{\textrm A}$ is the axial form factor and $F_{\textrm P}$ is the induced 
pseudoscalar form factor. The weak isovector Dirac and Pauli 
form factors, $F_1^{\textrm V}$ and
$F_2^{\textrm V}$, are related to the corresponding electromagnetic form
factors by the conservation of the vector current (CVC) hypothesis \cite{Walecka} plus, for NC reactions, a possible
isoscalar strange-quark contribution $F_i^{\mathrm s}$, i.e., 
\begin{eqnarray}
F_i^{\mathrm {V,p(n)}} &=& \left(\frac{1}{2} - 
2\sin^2{\theta_{\mathrm W}}\right)
 F_i^{\mathrm {p(n)}} -\frac{1}{2} F_i^{\mathrm {n(p)}} - 
 \frac{1}{2} F_i^{\mathrm s} \ 
 \ \ \ \ ({\mathrm {NC}}) \ , \nonumber \\
 F_i^{\mathrm V} &=& 
 F_i^{\mathrm p} - F_i^{\mathrm n} 
 \ \ \ \ ({\mathrm {CC}}) \ , 
\end{eqnarray}
where $\theta_{\mathrm W}$ is the Weinberg angle 
$(\sin^2{\theta_{\mathrm W}} \simeq 0.23143)$.
The electromagnetic form factors are taken from Ref. \cite{bba} and the strange
form factors are taken as \cite{alb}
\begin{eqnarray}
F_1^{\mathrm s}(Q^2) =  \frac {(\rho^{\mathrm s} + 
\mu^{\mathrm s}) \tau}{(1+\tau) (1+Q^2/M_{\mathrm V}^2)^2}\ , \ 
F_2^{\mathrm s}(Q^2) =  \frac {\left(\mu^{\mathrm s}-\tau \rho^{\mathrm s}  
\right)}{(1+\tau) (1+Q^2/M_{\mathrm V}^2)^2}\ ,
\label{eq.sform}
\end{eqnarray}
where $\tau = Q^2/(4M^2)$ and $M_{\mathrm V}$ =
0.843 GeV. The quantities $\mu_{\mathrm s}$ and $\rho_{\mathrm s}$ are related
to the strange magnetic moment and radius of the nucleus. 

The axial form factor is expressed as \cite{mmd}
\begin{eqnarray}
G_{\mathrm A} &=& \frac{1}{2} \left( \tau_3 g_{\mathrm A} - 
g^{\mathrm s}_{\mathrm A}\right) G\ \ \ \ ({\mathrm {NC}}) \ , \nonumber \\
G_{\mathrm A} &=&   g_{\mathrm A} G\ \ \ \ ({\mathrm {CC}}) \ , \label{eq.ga}
\end{eqnarray}
where $g_{\mathrm A} \simeq 1.26$, 
$g^{\mathrm s}_{\mathrm A}$ describes possible strange-quark contributions,  
$G = (1+Q^2/M_{\mathrm A}^2)^{-2}$, and $\tau_3 = +1 (-1)$ for proton (neutron)
knockout.
The axial mass has been taken 
from Ref. \cite{bernard} as $M_{\mathrm A}$ = (1.026$\pm$0.021) GeV, which is
the weighed average of the values obtained from (quasi-)elastic neutrino and
antineutrino scattering experiments.
The induced pseudoscalar form factor contributes only to CC scattering and can
be expressed as
\begin{equation}
F_{\textrm P}= \frac{2Mg_{\mathrm A}G}{m^2_{\pi} + Q^2} \ . \label{eq.formf}
\end{equation}

The differential cross section for the quasi-elastic 
$\nu$($\bar\nu$)-nucleus scattering is obtained from the contraction between 
the lepton and hadron tensors, as in Ref. \cite{Walecka}. 
After performing an integration over the solid angle of the 
final nucleon, we have
\begin{eqnarray}
\frac{\diff \sigma} {\diff \varepsilon \diff \Omega \diff {\mathrm {T_N}}} = 
 \frac{G_{{\mathrm F}}^2} {2 \pi^2} \, \varepsilon^2\cos^2 \frac {\vartheta}{2} 
  \Big [ v_0 R_{00} &+& v_{zz} R_{zz} - v_{0z} R_{0z} + v_T R_T \nonumber \\
  &\pm& v_{xy} R_{xy} \Big] \frac {|\p_{\mathrm N}| E _{\mathrm N}} 
 {(2 \pi)^3}\ .
\label{eq.cs}
\end{eqnarray} 
where $\vartheta$ is the lepton scattering angle and $E _{\mathrm N}$ the
relativistic energy of the outgoing nucleon. For CC processes 
$G_{{\mathrm F}}^2$ has to be multiplied by $\cos^2 \vartheta_{\mathrm C}$.
The coefficients $v$ are given as
\begin{eqnarray}
v_0 &=& 1 \ \ , \ \ 
v_{zz} = \frac {\omega^2} {|\q|^2}\ \ , \ \ 
v_{0z} = \frac{\omega} {|\q|}  \quad ,  \nonumber \\
v_T&=&\tan^2\frac {\vartheta}{2} + \frac{Q^2} {2|\q|^2} \ \ , \ \ 
v_{xy}= \tan \frac {\vartheta}{2} \left[ \tan^2\frac {\vartheta}{2} +
 \frac{Q^2} {|\q|^2} \right]^{\frac {1} {2}} \ \ , 
\label{eq.lepton}
\end{eqnarray}
where the mass of the final lepton has been neglected. If this is not the case,
as in CC scattering, the expressions for the coefficients $v$ can be found in
Ref. \cite{cc}.

The response functions $R$ are given in terms of the components of the 
hadron tensor as
\begin{eqnarray}
R_{00} &=& \int \diff \Omega_{\mathrm N}  \ W^{00}\ \ , \ 
R_{zz} = \int \diff \Omega_{\mathrm N}  \ W^{zz}\ \ , \ 
R_{0z} = \int \diff \Omega_{\mathrm N}  \ 2 \ \mathrm{Re} 
(W^{0z})\ , \nonumber \\
R_T  &=& \int \diff \Omega_{\mathrm N}  \ (W^{xx} + W^{yy})\ \ , \ 
R_{xy} =\int \diff \Omega_{\mathrm N}  \ 2\ 
\mathrm{Im}(W^{xy})\ .
\label{eq.rf}
\end{eqnarray}

The single differential cross section with respect to the outgoing 
nucleon kinetic energy $\mathrm {T_N}$ is obtained after performing an 
integration over the energy and the angle of the final lepton, i.e.,
\begin{eqnarray}
\frac {\diff \sigma} {\diff {\mathrm {T_N}}}  &=& \int 
\left( \frac{\diff \sigma} {\diff \varepsilon \diff \Omega \diff 
{\mathrm {T_N}}}
\right) \diff \varepsilon \diff \Omega  \ . \label{eq.cst} 
\end{eqnarray}
 
In the calculation of the transition amplitudes of Eq. \ref{eq.amp} the 
single-particle overlap functions $\varphi_n$ are taken as 
the Dirac-Hartree solutions of a relativistic Lagrangian,
containing scalar and vector potentials, obtained in the framework of the
relativistic mean field theory \cite{adfx,lala}. 
The relativistic single-particle scattering wave function 
is written as in Refs. \cite{meucci1,ee} in terms of its upper
component, following the direct Pauli reduction scheme, i.e.,
\inieq
 \chi_{\p_\mathrm{N}}^{(-)} =  \left(\begin{array}{c} 
{\displaystyle \Psi_{\textrm {f}+}} \\ 
\frac{\displaystyle 1} {\displaystyle 
M+ E +S^{\dagger}(E)-V^{\dagger}(E)}
{\displaystyle \mbox{\boldmath $\sigma$}\cdot\p
        \Psi_{\textrm {f}+}} \end{array}\right) \ ,
\fineq
where $S(E)$ and $V(E)$ are the scalar and vector 
energy-dependent
components of the relativistic optical potential for a nucleon
with energy $E$ \cite{chc}. 
The upper component, $\Psi_{\textrm {f}+}$, is related to a two-component
spinor, $\Phi_{\textrm{f}}$, which solves a
Schr\"odinger-like equation containing equivalent central and 
spin-orbit potentials, written in terms of the relativistic scalar and 
vector potentials \cite{clark,HPa}, i.e.,
\inieq
\Psi_{\textrm {f}+} = \sqrt{D^{\dagger}(E)}\ \Phi_{\textrm{f}} \ , 
\quad D(E) &=& 1 + \frac{S(E)-V(E)}{M + E} \ , 
\label{eq.darw}
\fineq
where $D(E)$ is the Darwin factor. 

We use in our calculations a relativistic optical potential with a real and 
an imaginary part which produces an absorption of flux. 
This is correct for an exclusive reaction, but would be incorrect for an 
inclusive one where the total flux must be conserved. In Refs. 
\cite{cc,ee,eepv} we presented an approach where FSI are treated in inclusive 
reactions by means of a complex optical potential and the total flux is 
conserved. In the present investigation, we consider situations where an 
emitted nucleon is always detected and treat the quasi-elastic neutrino 
scattering as a quasi-exclusive process, where the cross section is obtained 
from the sum of all the integrated exclusive one-nucleon knockout channels. 
Some of the reaction channels which are responsible for the imaginary 
part of the optical potential, like fragmentation of the nucleus, 
re-absorption, etc., are not included
in the experimental cross section when an emitted nucleon is detected. 
The outgoing nucleon, however, can be re-emitted after re-scattering 
in a detected channel, thus simulating the kinematics of a quasi-elastic 
reaction. 
The relevance of these contributions to the experimental cross section
depends on kinematics and should not be too large in the situations considered
in this paper. Anyhow, even if the use of an optical potential with an
absorptive imaginary part can introduce some uncertainties in the comparison 
with 
data, we deem it a more correct and clearer way to evaluate the effects of FSI. 
We note that the same uncertainties are also present in the analysis of 
exclusive quasi-elastic scattering, like $\left(e,e^{\prime}p\right)$, when 
the emission from deep states is considered.
An alternative treatment can be found in Ref. \cite{nieves}.

\section{Results and discussion}
\label{results}

%%%%%%%%%%%%%%%%%%%%%%%%%%%%%%%%%%%%%%%%%%%%%%%
\begin{figure}[h]
\begin{center}
\includegraphics[height=12cm, width=9cm]{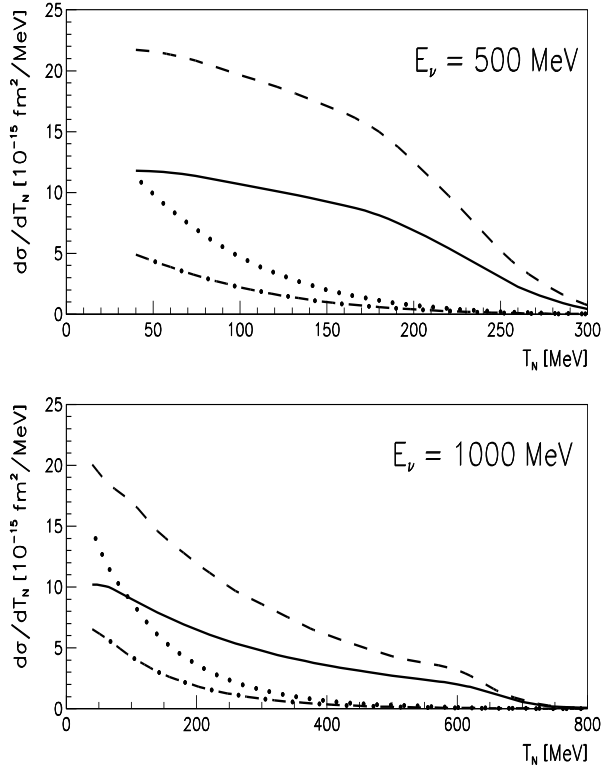} 
\vskip -0.7cm
\caption {Differential cross sections of the CC $\nu (\bar \nu)$ quasi-elastic
scattering on $^{12}$C
as a function of the outgoing nucleon kinetic energy
T$_{\mathrm N}$. Solid and dashed lines are the results in RDWIA and RPWIA,
respectively, for an incident neutrino. Dot-dashed and dotted lines are the 
results in RDWIA and RPWIA, respectively, for an incident antineutrino. 
}
\label{fig1}
\end{center}
\end{figure}
%%%%%%%%%%%%%%%%%%%%%%%%%%%%%%%%%%%%%%%%%%%%%%%
\begin{figure}[h]
\begin{center}
\includegraphics[height=12cm, width=9cm]{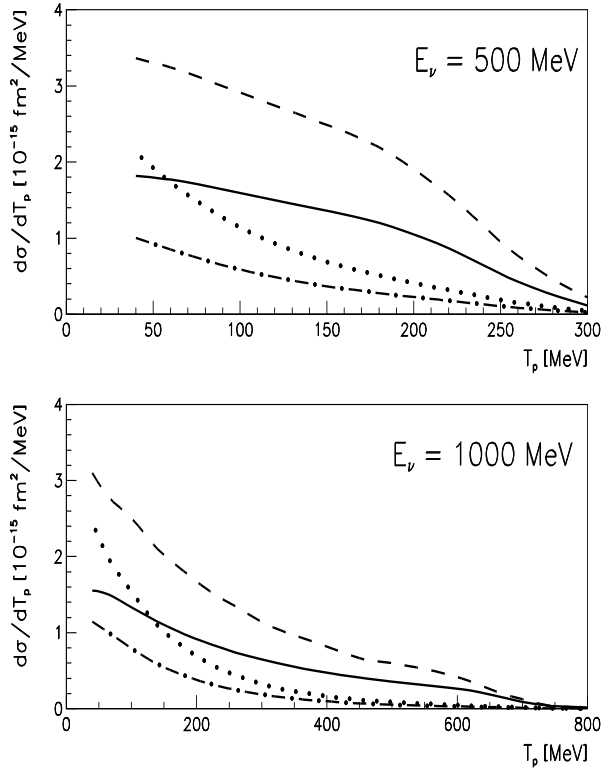} 
\vskip -0.7cm
\caption {Differential cross sections of the NC $\nu (\bar \nu)$ quasi-elastic
scattering on $^{12}$C
as a function of the outgoing proton kinetic energy
T$_{\mathrm p}$. Solid and dashed lines are the results in RDWIA and RPWIA,
respectively, for an incident neutrino. Dot-dashed and dotted lines are the 
results in RDWIA and RPWIA, respectively, for an incident antineutrino. The 
strangeness contribution is here neglected.}
\label{fig2}
\end{center}
\end{figure}
%%%%%%%%%%%%%%%%%%%%%%%%%%%%%%%%%%%%%%%%%%%%%%%
\begin{figure}[h]
\begin{center}
\includegraphics[height=12cm, width=9cm]{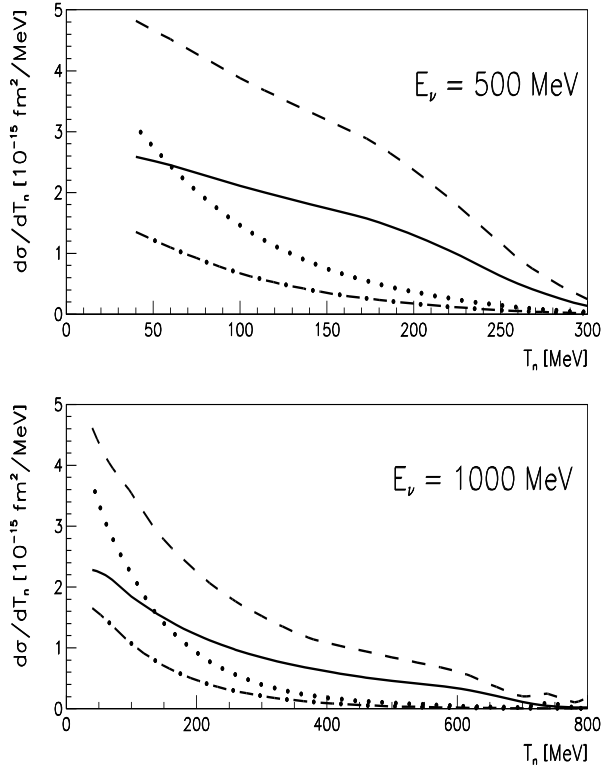} 
\vskip -0.7cm
\caption {The same as in Fig. \ref{fig2} but for neutron knockout.}
\label{fig3}
\end{center}
\end{figure}
%%%%%%%%%%%%%%%%%%%%%%%%%%%%%%%%%%%%%%%%%%%%%%%
\begin{figure}[h]
\begin{center}
\includegraphics[height=12cm, width=9cm]{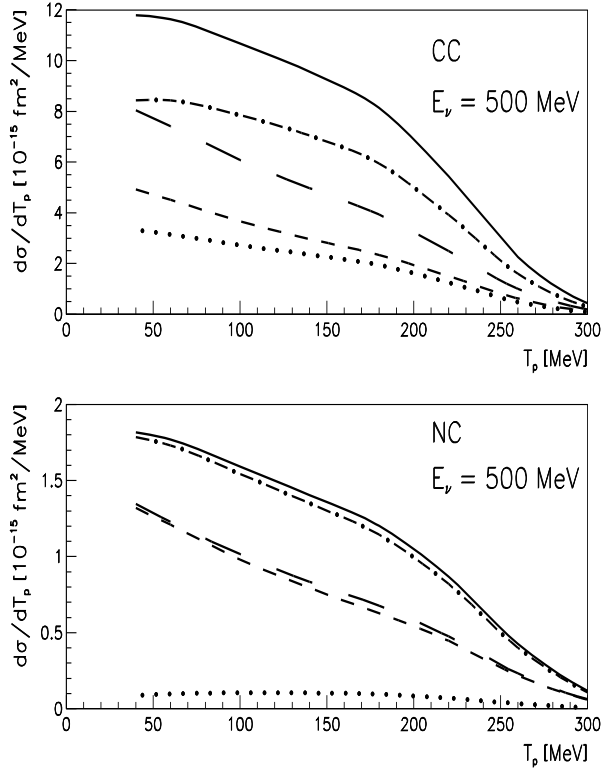} 
\vskip -0.7cm
\caption {Effect of the single-nucleon form factors on the CC and NC 
differential cross sections as a function of the outgoing proton kinetic energy.
Solid lines are the full results in RDWIA, dashed lines with 
$G_{\mathrm A} \neq 0$, $F_1 = F_2 = 0$, long dashed lines with $G_{\mathrm A} \neq 0$,
 $F_1 \neq 0$, and $F_2 = 0$, dot-dashed lines with $G_{\mathrm A} \neq
0$, $F_2 \neq 0$, and $F_1 = 0$, and dotted lines  
with $G_{\mathrm A} =0$, $F_1\neq 0$, and $F_2 \neq 0$. }
\label{form}
\end{center}
\end{figure}
%%%%%%%%%%%%%%%%%%%%%%%%%%%%%%%%%%%%%%%%%%%%%%%
\begin{figure}[h]
\begin{center}
\includegraphics[height=12cm, width=12cm]{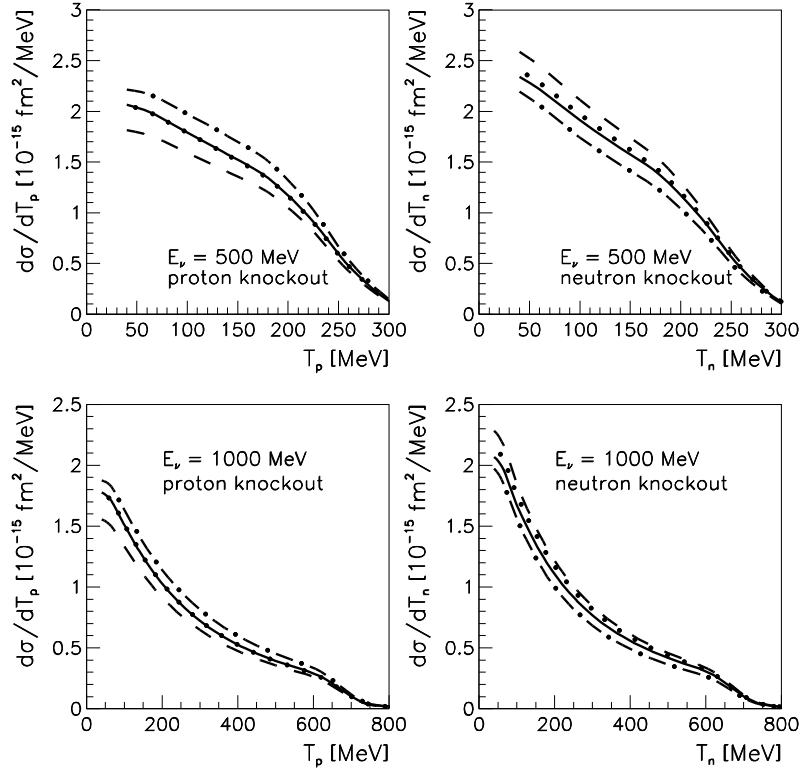} 
\vskip -0.7cm
\caption {Differential cross section of the NC $\nu$ quasi-elastic
scattering on $^{12}$C as a function of the outgoing nucleon 
kinetic energy. Dashed lines are the results with no 
strangeness contribution, 
solid lines with $g^{\mathrm s}_{\mathrm A} = -0.10$, dot-dashed lines
with $g^{\mathrm s}_{\mathrm A} = -0.10$ and $\mu^{\mathrm s} = -0.50$,
dotted lines with $g^{\mathrm s}_{\mathrm A} = -0.10$ and 
$\rho^{\mathrm s} = +2$.}
\label{fig4}
\end{center}
\end{figure}
%%%%%%%%%%%%%%%%%%%%%%%%%%%%%%%%%%%%%%%%%%%%%%%
%\begin{figure}[h]
%\begin{center}
%\includegraphics[height=12cm, width=9cm]{fig4} 
%\vskip -0.3cm
%\caption {Differential cross section of the NC $\nu$ quasi-elastic
%scattering on $^{12}$C as a function of the outgoing proton 
%kinetic energy $T_p$. Dashed lines are the results with no strangeness 
%contribution, 
%solid lines with $g^{\mathrm s}_{\mathrm A} = -0.10$, dot-dashed lines
%with $g^{\mathrm s}_{\mathrm A} = -0.10$ and $\mu^{\mathrm s} = -0.50$,
%dotted lines with $g^{\mathrm s}_{\mathrm A} = -0.10$ and 
%$\rho^{\mathrm s} = +2$.}
%\label{fig4}
%\end{center}
%\end{figure}
%%%%%%%%%%%%%%%%%%%%%%%%%%%%%%%%%%%%%%%%%%%%%%%
%\begin{figure}[h]
%\begin{center}
%\includegraphics[height=12cm, width=9cm]{fig5} 
%\vskip -0.3cm
%\caption {The same as in Fig. \ref{fig4} but for neutron knockout.}
%\label{fig5}
%\end{center}
%\end{figure}
%%%%%%%%%%%%%%%%%%%%%%%%%%%%%%%%%%%%%%%%%%%%%%%
\begin{figure}[h]
\begin{center}
\includegraphics[height=12cm, width=12cm]{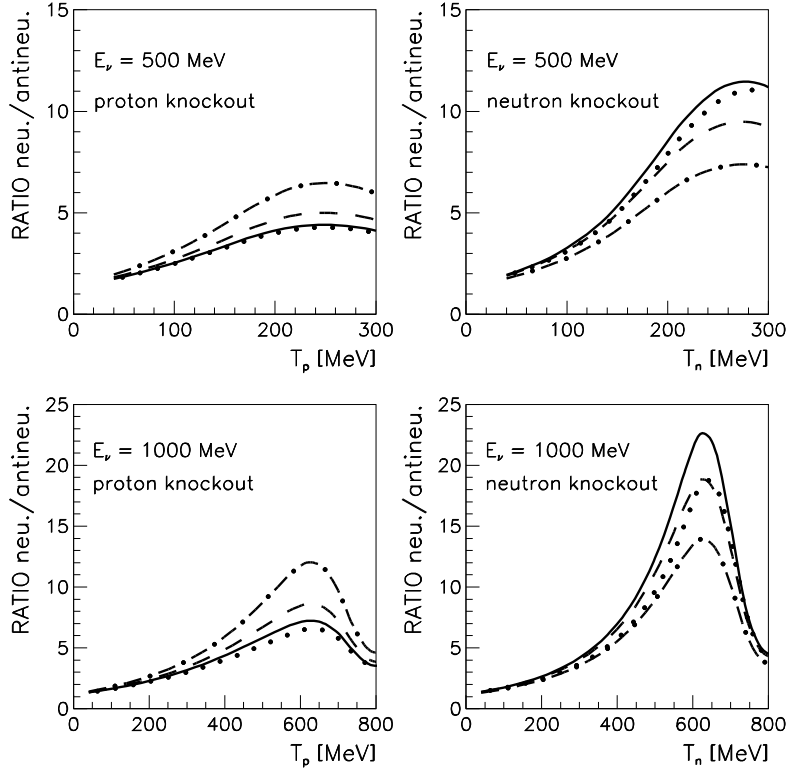} 
\vskip -0.7cm
\caption {Ratio of neutrino-to-antineutrino NC cross sections 
on $^{12}$C as a function of the outgoing nucleon kinetic energy.
Line convention as in Fig. \ref{fig4}.}
\label{fig6}
\end{center}
\end{figure}
%%%%%%%%%%%%%%%%%%%%%%%%%%%%%%%%%%%%%%%%%%%%%%%
%\begin{figure}[h]
%\begin{center}
%\includegraphics[height=12cm, width=9cm]{fig7} 
%\vskip -0.3cm
%\caption {Upper panel: differential cross sections of the $\nu (\bar \nu)$}
%\label{fig7}
%\end{center}
%\end{figure}
%%%%%%%%%%%%%%%%%%%%%%%%%%%%%%%%%%%%%%%%%%%%%%%
\begin{figure}[h]
\begin{center}
\includegraphics[height=12cm, width=12cm]{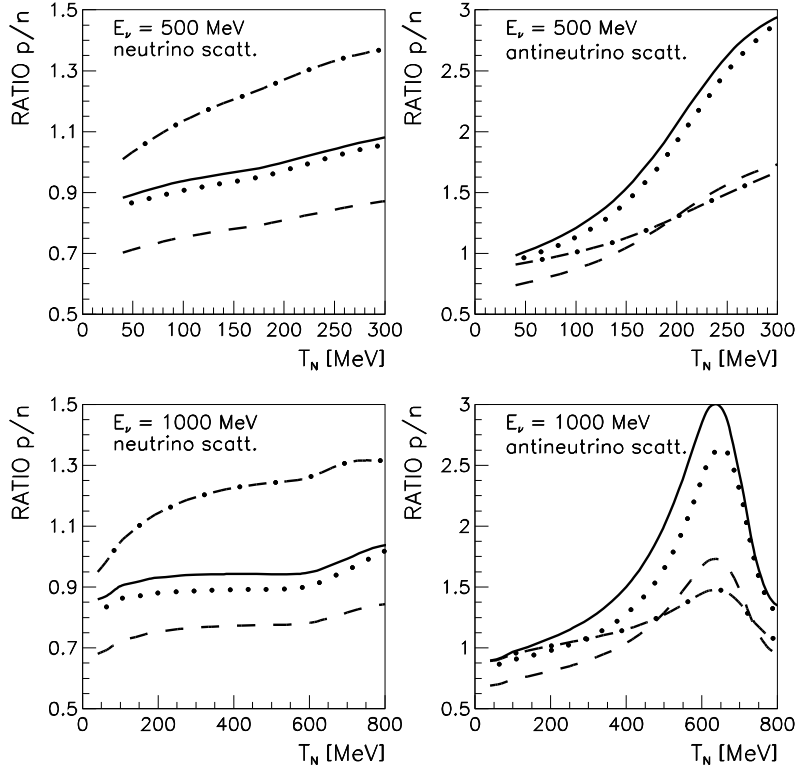} 
\vskip -0.7cm
\caption {Ratio of proton-to-neutron NC cross sections of the $\nu$ $(\bar\nu)$
quasi-elastic scattering on $^{12}$C as a function of 
T$_{\mathrm N}$. Line convention as in Fig. \ref{fig4}.}
\label{fig8}
\end{center}
\end{figure}
%%%%%%%%%%%%%%%%%%%%%%%%%%%%%%%%%%%%%%%%%%%%%%%
\begin{figure}[h]
\begin{center}
\includegraphics[height=12cm, width=12cm]{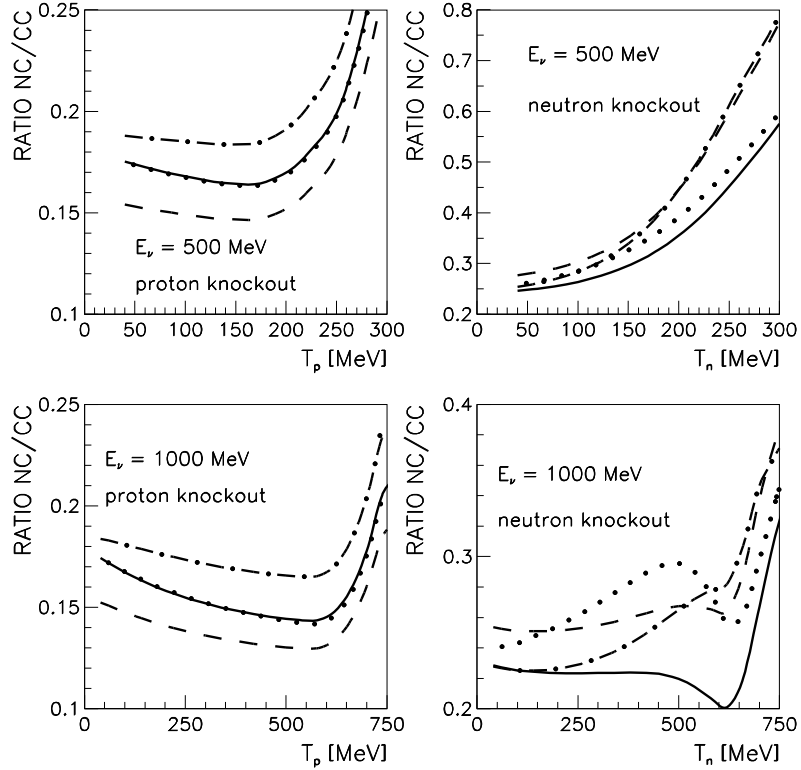} 
\vskip -0.7cm
\caption {Ratio of neutral-to-charged current cross sections of the $\nu$ $
(\bar\nu)$
quasi-elastic scattering on $^{12}$C as a function of the
outgoing nucleon kinetic energy. Line convention as in Fig. \ref{fig4}.}
\label{fig9}
\end{center}
\end{figure}
%%%%%%%%%%%%%%%%%%%%%%%%%%%%%%%%%%%%%%%%%%%%%%%
%\begin{figure}[h]
%\begin{center}
%\includegraphics[height=12cm, width=9cm]{bnl2} 
%\vskip -0.3cm
%\caption {Upper panel: differential cross sections of the $\nu (\bar \nu)$}
%\label{fig10}
%\end{center}
%\end{figure}
%%%%%%%%%%%%%%%%%%%%%%%%%%%%%%%%%%%%%%%%%%%%%%%
\begin{figure}[h]
\begin{center}
\includegraphics[height=12cm, width=9cm]{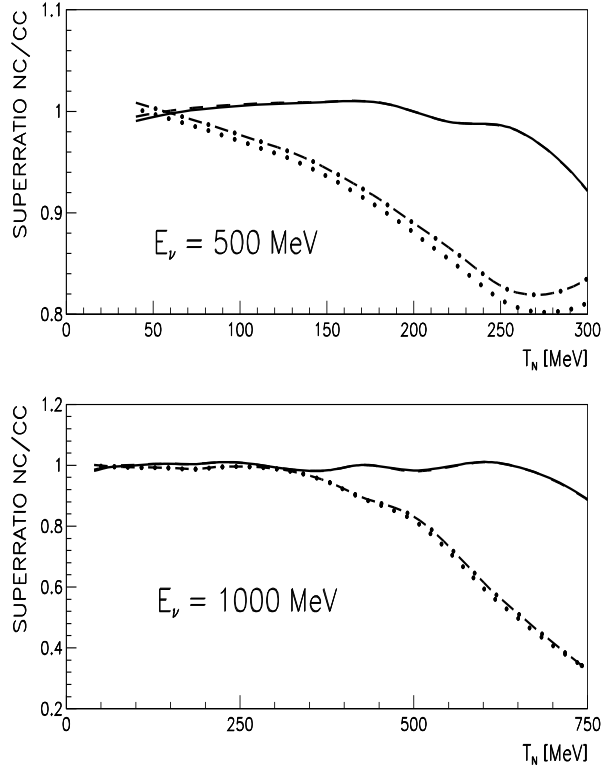} 
\vskip -0.7cm
\caption {Ratio of
(NC/CC)$_{\mathrm {RDWIA}}$-to-(NC/CC)$_{\mathrm {RPWIA}}$ cross sections as
a function of the outgoing nucleon kinetic energy T$_{\mathrm N}$. Dashed and
solid lines are the results with no 
strangeness contribution and with $g^{\mathrm s}_{\mathrm A} = -0.10$,
respectively, for an incident neutrino and proton knockout. Dot-dashed and
dotted lines are the results with no 
strangeness contribution and with $g^{\mathrm s}_{\mathrm A} = -0.10$,
respectively, for an incident antineutrino and neutron knockout.
}
\label{fig11}
\end{center}
\end{figure}
%%%%%%%%%%%%%%%%%%%%%%%%%%%%%%%%%%%%%%%%%%%%%%%
\begin{figure}[h]
\begin{center}
\includegraphics[height=12cm, width=9cm]{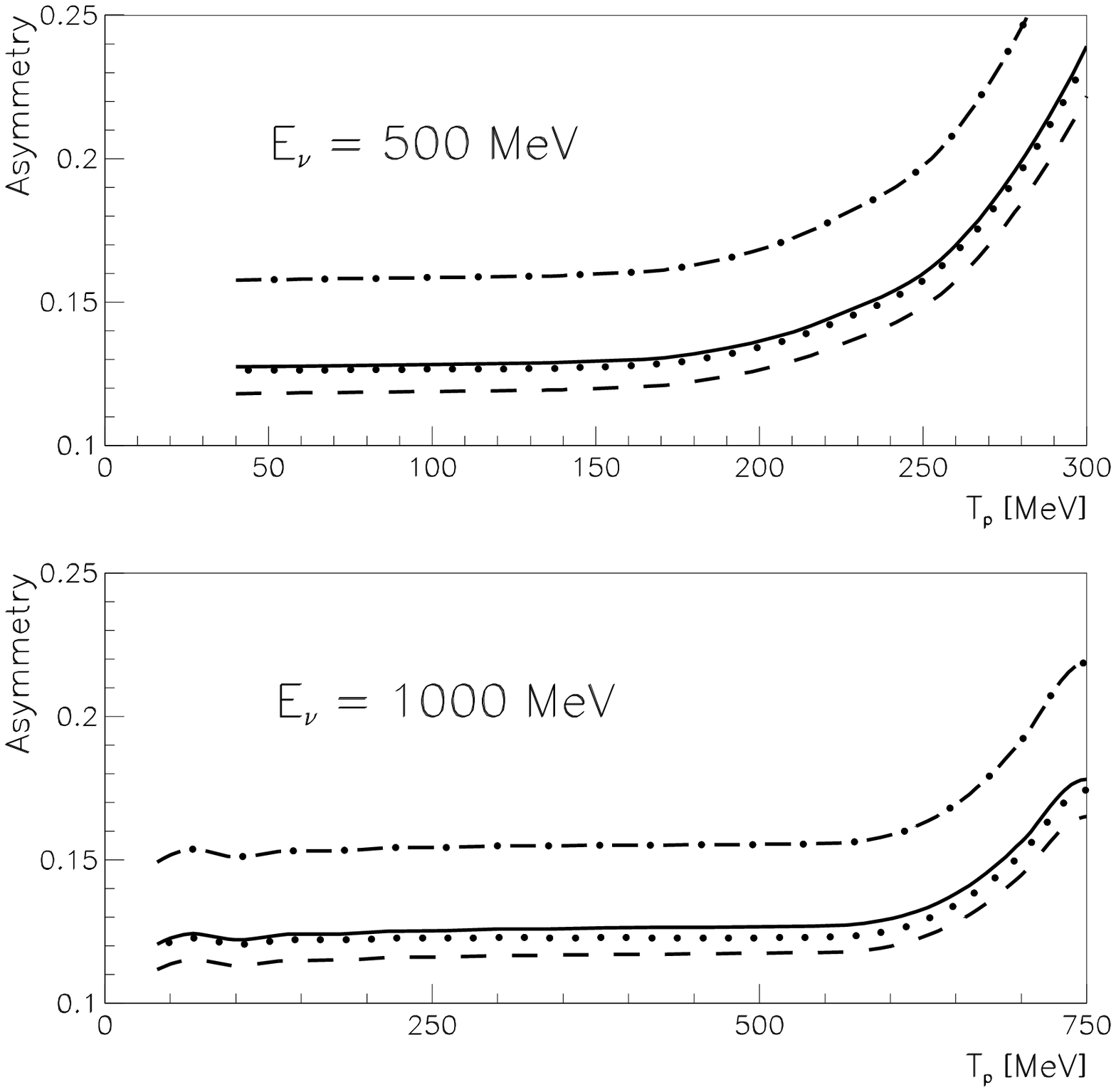} 
\vskip -0.7cm
\caption {The integral asymmetry of Eq. \ref{eq.asy} for $\nu$ quasi-elastic
scattering on $^{12}$C as a function of the outgoing nucleon 
kinetic energy T$_{\mathrm N}$. Line convention as in Fig. \ref{fig4}.}
\label{fig12}
\end{center}
\end{figure}
%%%%%%%%%%%%%%%%%%%%%%%%%%%%%%%%%%%%%%%%%%%%%%%

Results are presented for NC and CC neutrino and antineutrino scattering from 
$^{12}$C in an energy range up to 1000 MeV, where the quasi-elastic one-nucleon 
knockout is expected to be the most important contribution to the cross 
section. The main aim of our investigation is to study the effects of FSI and 
of a non-zero contribution of the strangeness to the form factors. 

In the calculations we have used the same relativistic bound state wave 
functions and optical potentials as in
Refs.~\cite{meucci1,meucci2,ee,eepv}, where the RDWIA was able to 
reproduce $\left(e,e^{\prime}p\right)$, $\left(\gamma,p\right)$, and 
$\left(e,e^{\prime}\right)$ data. 
The relativistic bound state wave functions have been obtained from 
Ref.~\cite{adfx}, where relativistic Hartree-Bogoliubov equations are solved in
the context of a relativistic mean field theory and reproduce
single-particle properties of several spherical and deformed nuclei~\cite{lala}. 
Calculations performed with different parameterizations for the bound state wave
functions give only negligible differences with respect to the results shown in
this paper. 

The scattering states are computed by means of
the energy-dependent and A-dependent EDAD1 complex phenomenological optical 
potential of Ref. \cite{chc}, which includes the Coulomb interaction and is 
fitted to proton elastic scattering data on several nuclei in an energy range 
up to 1040 MeV. The neutron scattering states are computed with the same 
optical potential but neglecting the Coulomb interaction. Calculations
performed with the EDAD2 optical potential \cite{chc} do not change
significantly our results.

The initial states $\varphi_n$ are single-particle one-hole states in the target
with an unitary spectral strength. The sum  in Eq. \ref{eq.ha1} runs over all 
the occupied states in the shell model.
In this way we include the contributions of all the nucleons in the nucleus, 
but disregard effects of correlations. These effects, however, are expected to 
be small on the semi-exclusive cross section and, moreover, should not affect
the role of FSI and of the strange-quark content of the form factors, which are
the aims of the present investigation.

In order to study the effects of FSI, we first compare results  of the CC and NC
$\nu$($\bar\nu$)-nucleus cross section in RPWIA and RDWIA. In Fig. \ref{fig1}, 
the CC cross section for the $^{12}$C$\left(\nu_{\mu},\mu^-p\right)$ and 
$^{12}$C$\left(\bar\nu_{\mu},\mu^+n\right)$ reactions are presented at 
$E_{\nu}= 500$ and $1000$ MeV. FSI effects are large and reduce the cross 
section of $\simeq 50$\%. This reduction is in agreement with the results 
obtained in the electromagnetic one-nucleon knockout. 
%We have performed calculations with different
%parameterizations for the bound state wave-functions and found almost negligible
%differences. We remind that for the case of the inclusive scattering this
%approach, that is based on the sum of all the integrated exclusive reaction
%channels does not conserve the flux and therefore it is not correct. 
%In Refs. \cite{cc,ee,eepv} we have
%presented and discussed a simple way to include FSI preserving flux conservation
%for inclusive reactions.
%The results for the reaction 
%$^{12}$C$\left(\bar\nu_{\mu},\mu^+n\right)$ are also shown in Fig. \ref{fig1}. 

In Figs. \ref{fig2} and \ref{fig3} the NC $\nu$($\bar\nu$)-nucleus cross 
sections are displayed as a function of the emitted nucleon kinetic energy 
for proton and neutron knockout, respectively. Also for the NC
scattering the RDWIA results are reduced up to $\simeq 50$\% with respect to 
the RPWIA ones. We note that the cross sections in Figs. \ref{fig2} and 
\ref{fig3}  are different from our previous results of Ref.~\cite{nc}.
The calculations presented in this paper have been obtained after a careful 
check, where some inconsistencies of our previous calculations were found and 
eliminated. These new results are consistent with those of 
Ref.~\cite{Martinez:2005xe} and are larger than those of Ref.~\cite{vdv}.

The contribution of the single-nucleon form factors to the NC and CC cross 
sections is investigated in Fig. \ref{form} for proton knockout at 
$E_{\nu}= 500$ MeV. No single form factor reproduces the full cross section. 
In the case of CC scattering, the major contributions come from 
calculations with only $G_{\mathrm A}$ and $F_1$ or $G_{\mathrm A}$ and $F_2$.
These contributions are however $\simeq 30$\% lower than the full result.
This indicates that all form factors and their interference terms are 
important for the CC reaction. In contrast, in the NC process the weak Dirac
form factor $F_1$ is suppressed by the mixing angle term 
$\left(1/2 - 2\sin^2{\theta_{\mathrm W}}\right) \simeq 0.04$. The cross 
sections obtained whith only $G_{\mathrm A}$ and $F_2$ are almost 
identical to the full results. 
The axial-vector form factor plays a dominant role in the NC reaction; in fact, 
when it is neglected, the cross section becomes very small.

The effects of a non-zero strange-quark contribution to the axial-vector and to
the vector form factors on the NC cross sections are shown in
Fig. \ref{fig4} %-\ref{fig5} 
both for proton and neutron emission. A precise determination of the exact 
values of $g^{\mathrm s}_{\mathrm A}$, 
$\mu^{\mathrm s}$, and $\rho^{\mathrm s}$ is beyond the scope of this paper. 
Thus, here we have chosen \lq\lq typical\rq\rq\ values for the strangeness 
parameters to show up their effect on the results, being aware that the value 
of $g^{\mathrm s}_{\mathrm A}$ is correlated to the value of the axial mass
$M_{\mathrm A}$ and that the values of $\mu^{\mathrm s}$ and $\rho^{\mathrm s}$ 
are also highly correlated
(see, e.g., Ref.~\cite{hor}). In our calculations we have used 
$g^{\mathrm s}_{\mathrm A} = -0.10$, $\mu^{\mathrm s} = -0.50$, and 
$\rho^{\mathrm s} = +2$. The opposite sign for $\mu^{\mathrm s}$ and 
$\rho^{\mathrm s}$ agrees with first results from HAPPEX \cite{aniol}. The
results with $g^{\mathrm s}_{\mathrm A} = -0.10$ are enhanced in the case of 
proton knockout and reduced in the case of neutron knockout by $\simeq 10$\% 
with respect to those with $g^{\mathrm s}_{\mathrm A} = 0$ . 
The effect of $\mu^{\mathrm s}$ is
large and comparable to that of $g^{\mathrm s}_{\mathrm A}$, whereas the
contribution of $\rho^{\mathrm s}$ is very small for neutron knockout and
practically negligible for proton knockout.  

The role of the strangeness contribution can also be studied in the ratio of
neutrino-to-antineutrino induced NC cross sections. In Fig. \ref{fig6}
%-\ref{fig7}
our RDWIA results are displayed as a function of T$_{\mathrm N}$ for proton and
neutron knockout. This ratio is
very sensitive to $\Delta {\mathrm s}$ and presents a maximum at T$_{\mathrm N} \simeq
0.6$ E$_{\nu}$. In the case of proton knockout the ratio is reduced by 
$g^{\mathrm s}_{\mathrm A}$ and enhanced by $\mu^{\mathrm s}$. 
In contrast, for neutron knockout  the ratio is enhanced
by $g^{\mathrm s}_{\mathrm A}$ and reduced by $\mu^{\mathrm s}$. The strange
radius $\rho^{\mathrm s}$ reduces the ratio. This effect is very small
for proton knockout and larger for neutron knockout and increases with the
incident neutrino and antineutrino energy.

Another interesting quantity proposed to
study strangeness effects is the ratio of proton-to-neutron (p/n) NC 
cross sections \cite{alberico,hor,gar92}. This ratio is 
very sensitive to the strange-quark contribution as the axial-vector
strangeness $g^{\mathrm s}_{\mathrm A}$ interferes with the isovector
contribution $g_{\mathrm A}$ with one sign in the 
numerator and with the opposite sign in the denominator (see Eq. \ref{eq.ga}). 
Moreover, it is expected to be less sensitive to distortion effects. 
The p/n ratio calculated in RDWIA for an incident neutrino or antineutrino 
is displayed in 
Fig. \ref{fig8} as a function of T$_{\mathrm N}$. The RPWIA results are almost 
coincident (up to a few percent) and are not shown in the figure. 
The p/n ratio for an incident neutrino is enhanced by a factor 
$\simeq 20-30$\% when 
$g^{\mathrm s}_{\mathrm A}$ is included and by $\simeq 50$\% when both  
$g^{\mathrm s}_{\mathrm A}$ and $\mu^{\mathrm s}$ are included. 
A minor effect is produced also in this case by $\rho^{\mathrm s}$, which gives 
only a slight reduction of the p/n ratio. 
The results for an incident antineutrino are quite different. The ratio is
largely enhanced when $g^{\mathrm s}_{\mathrm A}$ is included but the
enhancement is canceled when also $\mu^{\mathrm s}$ is considered. Also in this
case $\rho^{\mathrm s}$ plays a minor role.
Precise measurements of the p/n 
ratio appear however problematic due to the difficulties associated with 
neutron detection. 
This is the reason why
the most attractive quantity to extract experimental information about
$\Delta {\mathrm s}$ seems the ratio of the neutral-to-charged (NC/CC) cross 
sections. 
In fact, although sensitive to $\Delta {\mathrm s}$ only in the numerator, the 
NC/CC ratio is simply related to the number of events with an outgoing
proton and a missing mass with respect to the events with an outgoing proton in
coincidence with a muon. 

Our RDWIA results for the NC/CC ratio are presented in Fig. \ref{fig9} 
%-\ref{fig10} 
as a function of T$_{\mathrm N}$ for proton and
neutron emission. For the case of neutron knockout,
the incident particle is supposed to be an antineutrino.  
The results for proton knockout show similar features at different 
energies of the incident neutrino. 
The fact that the CC cross section goes to zero more rapidly
than the corresponding NC one (because of the muon mass) causes the 
enhancement of the ratio at large values of T$_{\mathrm p}$. The inclusion of a
strangeness contribution produces a somewhat constant enhancement of the results
with respect to the case $\Delta {\mathrm s} = 0$. The simultaneous inclusion of 
$g^{\mathrm s}_{\mathrm A}$ and $\mu^{\mathrm s}$ gives an enhancement that is 
about a factor of 2 larger than the one corresponding to the case with
only $g^{\mathrm s}_{\mathrm A}$ included. The effect of $\rho^{\mathrm s}$ is
very small. The results for an incident antineutrino and neutron knockout 
appear quite different at different energies of the incident antineutrino. 
Moreover, the effects of
strangeness are somewhat energy dependent. In this case the larger
effect, i.e., a reduction of the ratio, is obtained when only 
$g^{\mathrm s}_{\mathrm A}$ is included. The global effect is reduced when also 
$\mu^{\mathrm s}$ is considered. 

Ratios of cross
sections are attractive quantities because they are supposed to be rather
insensitive to FSI between the outgoing nucleon and the residual nucleus.
In order to investigate this point we show in Fig. \ref{fig11} our results for 
the ratio of (NC/CC)$_{\mathrm {RDWIA}}$-to-(NC/CC)$_{\mathrm {RPWIA}}$. 
Results for proton knockout are always close to unity, apart from some mild 
oscillations, up to a few percent. In the case of an incident antineutrino and neutron knockout, the results
are lower than unity. This is due both to Coulomb distortion and to the
different coupling of the optical potential with proton and neutron currents, 
and means that FSI cannot be neglected in the determination of the NC/CC ratio
with an incident antineutrino and neutron knockout. The effects of the
axial-vector strangeness are also shown in Fig. \ref{fig11}. The results do not
show significant differences with respect to the $\Delta {\mathrm s} = 0$ case.

In order to show up the effect of the strange-quark contribution, in Ref.
\cite{alberico} it is proposed to study the asymmetry
\begin{eqnarray}
\mathcal A = \frac{\left[\left(\diff\sigma/\diff\mathrm {T_N}\right)_{\nu}
- \left(\diff\sigma/\diff\mathrm {T_N}\right)_{\bar\nu}\right]^{\mathrm {NC}}}
{\left[\left(\diff\sigma/\diff\mathrm {T_N}\right)_{\nu}
- \left(\diff\sigma/\diff\mathrm {T_N}\right)_{\bar\nu}\right]^{\mathrm {CC}}}
\  .
\label{eq.asy}
\end{eqnarray} 
In Fig. \ref{fig12} we show our results corresponding to the emission of one 
proton in NC scattering as a function of T$_{\mathrm p}$. The sensitivity to 
the strange-quark content is visible when $g^{\mathrm s}_{\mathrm A}$ and 
$\mu^{\mathrm s}$ are included. The role of $\rho^{\mathrm s}$ is also in this
case less
significant. These results are in agreement with those of Ref. \cite{alberico}.
   
\section{Summary and conclusions}
\label{conc}

We have presented relativistic calculations for charged- and neutral-current 
$\nu$($\bar\nu$)-nucleus quasi-elastic scattering. The reaction mechanism is
assumed to be a direct one, i.e., the incident neutrino (antineutrino) 
interacts with
only one nucleon in the target nucleus and the other nucleons behave as 
spectators. A sum over all single particle
occupied states is performed, using an independent particle model to describe 
the structure of the nucleus. 
The scattering state is an optical-model wave function. 
Results for the $^{12}$C target nucleus have been
presented at neutrino (antineutrino) energies up to 1000 MeV. 

FSI are an important ingredient of the calculations, as the optical
potential produces a large reduction of the cross sections and gives a slightly
different effect on proton and neutron emission. 
The sensitivity to
the strange-quark content of the form factors has been investigated. 
The cross section is increased by strange-quark contribution
for proton knockout and decreased for neutron knockout. 
The ratio of neutrino-to-antineutrino cross sections is very sensitive to
$\Delta {\mathrm s}$.
An enhancement of the ratio of proton-to-neutron cross sections is 
obtained which is almost proportional to $g^{\mathrm s}_{\mathrm A}$. 
The enhancement is almost independent of FSI. 
The sensitivity to the strange-quark 
contribution of the vector form factors has also been discussed: 
$\mu^{\mathrm s}$ enhances the p/n ratio while $\rho^{\mathrm s}$ gives only 
small effects. 
An attractive quantity for studying the sensitivity to strange-quark effects 
is the ratio of neutral-to-charged current cross sections. With $g^{\mathrm
s}_{\mathrm A} = -0.10$ an enhancement of the NC/CC ratio
of $\simeq 15$\% relative to the case $\Delta {\mathrm s} = 0$ is obtained. The
inclusion of $\mu^{\mathrm s} = -0.50$ produces a further increase of 
$\simeq 30$\%.
The effects of FSI are small in the case of proton knockout but cannot be
neglected in the case of neutron knockout.

\begin{acknowledgments}

We would like to thank Prof. Stephen Pate for useful discussions and comments.

\end{acknowledgments}

%%%%%%%%%%%%%%%%%%%%%%%%%%%%%%%%%%%%%%%%%%%%%%%

%\vskip 4cm

\end{document}